# Single photoelectron spin detection and angular momentum transfer in a gate defined quantum dot

Takafumi Fujita[1], Kazuhiro Morimoto[1], Haruki Kiyama[1], Giles Allison[1,2], Marcus Larsson[1], Arne Ludwig[3], Sascha R. Valentin[3], Andreas D. Wieck[3], Akira Oiwa[1,4] & Seigo Tarucha[1,2]

[1]Department of Applied Physics, The University of Tokyo, 7-3-1 Hongo, Bunkyo-ku, Tokyo 113-8656, Japan
[2]Center for Emergent Matter Science (CEMS), RIKEN, 2-1 Hirosawa, Wako-shi, Saitama 351-0198, Japan
[3]Lehrstuhl für Angewandte Festkörperphysik, Ruhr-Universität Bochum, Universitätsstraße 150, Gebäude NB, D-44780 Bochum, Germany
[4]Institute of Scientific and Industrial Research, Osaka University, 8-1 Mihogaoka, Ibaraki, Osaka 567-0047, Japan

Recent innovations in fabricating nanoscale confined spin systems have enabled investigation of fundamental quantum correlations between single quanta of photons and matter states [1-5]. Realization of quantum state transfer from photon polarization to electron spin using gate defined quantum dots (QDs) may give evidence of preserved coherence of angular momentum basis states at the photon-spin interface. The interface would enlarge the concept of quantum information technology, in which single photogenerated electron spins are manipulated with the dots [6-10], but this remains a serious challenge [11-19]. Here, we report the detection of single electron spins generated by polarized single photons via a double QD (DQD) to verify the angular momentum transfer from single photons to single electrons. Pauli spin blockade (PSB) is used to project the photoelectron spin state onto the up or down spin state. Our result promises the realization of coherent quantum state transfer and development of hybrid photon and spin quantum technology.

The quantum interface between single photons and different two level systems, initially established for atoms and ions [20,21] and recently excitons [22], electrons [1,4,5] and superconducting circuits [23] has been attracting massive interest. These studies exemplify the concepts of quantum coherence and entanglement between hetero-particles that have not yet been well established for solid-state systems. In addition they may provide necessary elements for hybrid photon-spin quantum information systems including quantum repeaters for long-distant quantum communication and quantum nodes for quantum networks [11].

Quantum electrodynamics has been intensively studied between micro-wave photons and superconducting circuits or QD charge states that are established via the photon number state [24]. On the other hand, the photon polarization state is a more useful quantum number to couple with spins. The photon-spin interface has been studied for charged exciton states in self-assembled QDs and electron spin states in nitrogen-vacancy centres [1,4,5]. However, these systems have only been accessible by optical means to date. Electrical manipulation is far more developed in gate defined QDs, therefore they are ideal candidates for implementing the quantum interface by establishing optical access to their spin states.

Electron spins in gate defined QDs have been investigated for application to quantum information technologies owing to their long spin coherence time, tunability and scalability [6-10]. In particular, the electrically feasible manipulation and detection of entangled two electron spin states [10] has motivated the implementation of gate defined QDs as a quantum repeater. Following the theoretical proposal [12], quantum state transfer has been experimentally realized in a GaAs based double heterojunction quantum well (QW) from an ensemble of polarized photons to an ensemble of electron spins [13,14]. Efforts to deal with single quanta as opposed to the ensembles have been made using QDs but so far has only been successful for the charge state [15-19].

Single electron spins are measured using combined techniques of spin-to-charge information conversion with an appropriate spin effect such as the Zeeman effect and PSB [6,7]. Dynamical measurements enable studies on the spin evolution via hyperfine interaction with the nuclear spins and spin-orbit interaction [25-27]. The measurement of inter-dot charge tunnelling in a DQD can serve as a useful signature to observe the spin dependent tunnelling due to PSB in combination with the photoelectron trapping.

In this work, we verify the angular momentum transfer from single photons to single electron spins in a DQD based on a GaAs QW. We confirm that PSB can be used to detect two electron spin configurations in real-time measurements, and using this scheme, we detect the single electron spins that are created by circularly polarized photons and excited from the heavy hole band (Fig. 1a).

The DQD is laterally formed by gating a two-dimensional electron gas accumulated in a GaAs QW. Laser pulses are irradiated within the charge detection rise time through the aperture of a metal mask opened above the left QD (Fig. 1b). The device is mounted in a dilution refrigerator (base temperature 25 mK) equipped with a superconducting magnet, where the magnetic field is aligned perpendicular to the QW plane.

The whole procedure of detecting single photoelectron spins using PSB is explained in the following initialization, excitation, and measurement steps (Fig. 1c). An electron spin in the DQD is initialized to a (0,1) charge state, where the left (right) number is the number of electrons in the left (right) dot, to determine the projection axis of the photon angular momentum in an external magnetic field. Upon photoexcitation, an electron-hole pair can be excited, of which only the electron is trapped in the DQD potential, resulting in a (1,1) two electron spin state. When the PSB condition is met and the energy levels of the (1,1) and (0,2) states are aligned, the charge state will be locked in the (1,1) state if it is a (1,1) spin triplet state, but not if it is a (1,1) spin singlet state. We utilize a condition where the inter-dot tunnelling time is long compared to the $S$ and $T_0$ mixing by the hyperfine interaction, leading these anti-parallel spin states to couple equally to the (0,2) singlet state [27]. As a consequence, the excitation of an anti-parallel spin state is followed by an inter-dot electron tunnelling or charge movement between (1,1) and (0,2) states which is detected in real-time with a charge sensor.

The appropriate DQD energy level for the spin detection scheme is identified near the (1,1) and (0,2) inter-dot charge transition region (Fig. 2a). The inter-dot tunnelling rate conditions and photoelectron trapping signals in an external magnetic field $B = 0$ are similar to those presented in Ref. 18, but in between (1,0) and (0,1) charge states (see specific details for this sample in the Supplementary material). Here we show results for the different electron number in a finite magnetic field where we satisfy the PSB condition.

Figure 2b shows a typical charge sensor current $I_{sensor}$ at $B = 1$ T where we observe the PSB feature on the inter-dot transition line of (1,1) and (0,2) states (point ● in Fig. 2a). We clearly distinguish the time blocked in the (1,1) state and the time tunnelling back and forth between the (1,1) and (0,2) states. The difference between these spin states is quantified from the double exponential decay in the histogram of the (1,1) state residing time. For example Fig. 2c indicates the distinguishability from parallel spin PSB to anti-parallel spin inter-dot tunnelling with the corresponding decay times $\tau_{slow} = 11.3$ ms and $\tau_{fast} = 496$ µs, respectively. The contrast of the two decay times is kept large in the presence of a magnetic field sufficiently larger than where the hyperfine interaction limits the PSB lifetime, and lower than where the parallel spin state directly tunnels to the higher orbital state in $T_0(0,2)$ state (400 mT < $B$ < 2 T).

To combine this spin configuration detection with photoelectron trapping, the dot is initialized to the (0,1) state (point ★ in Fig. 2a) along the red dashed line with the (1,1) and (0,2) excited state levels aligned. When the PSB condition is met, single-shot traces show photoelectrons blocked at the (1,1) state (Fig. 2d). Similar type of data but at different magnetic fields are shown to illustrate the external magnetic field dependence of the PSB configuration of photogenerated parallel spin states. The increasing dependence of the blocked (1,1) state lifetime against the magnetic field can be quantitatively explained by the nuclear spin field fluctuation [27] and is discussed in the Supplementary material. Note that because linearly polarized photons are used, we obtain a different type of data set showing inter-dot electron tunnelling upon photoexcitation due to formation of anti-parallel spin states with a similar probability. For the stability of the measurement, the photon flux is adjusted to observe one photoelectron trapping event on the left QD per approximately 10 shots.

One of the required conditions to ensure the polarization-to-spin transfer is to resonantly excite the heavy hole band. The resonance is observed as a detection probability peak in the incident photon energy dependence (Fig. 2e). Another requirement is to efficiently polarize the reference spin by increasing the external magnetic field. We worked at a higher field of 1.65 T to increase the spin splitting compared to the thermal energy. Figure 3a shows examples of the typical single photoelectron spins measured as up (Fig. 3a top) or down (Fig. 3b bottom) in the above conditions. The difference in the decay times $\tau_{slow} \sim 500$ ms and $\tau_{fast} \sim 20$ ms is large enough to distinguish between the two types of signals with an accuracy over 90 % (See Supplementary material for more discussions on the detection accuracy).

With these preparations the single photon angular momentum are detected. Figure 3b plots the probability of detecting a spin towards the magnetic field direction (upper trace of Fig. 3a). Over a thousand single-shot data are collected for each data point with polarization varied from $\sigma^-$ to $\sigma^+$ via linear and opposite directions of the $B$ field. In the positive field, the $\sigma^-$ irradiation gives a maximum probability of finding an up spin, while the $\sigma^+$ irradiation gives the minimum probability because down spin is predominantly created. This polarization-to-spin relation corresponds to the selection rules of the heavy hole band excitation which infers the circular polarization to electron spin conversion. The rest of the plots are expected to follow a sine curve against the wave plate rotation angle, if the photon Stokes vector is mapped on the Bloch vector and projected on a basis defined by the external magnetic field. We observe the one to one correspondence for the measured points. By reversing the field, a mirrored trace of the detection probabilities is obtained as further justification of the spin transfer. Here, we have verified in a single-shot manner that the angular momentum of a single circular polarized photon is transferred to the electron spin.

The probability of detecting an up or down spin with circular polarized photon is currently limited to ~60 % in our experiment. This can be improved to over 90 % by improving the electron spin readout scheme by having a larger contrast of tunnelling times between the two spin states and avoiding thermal broadening (details on the amplitude and the slight offset with respect to 50 % are discussed in the Supplementary material). Our detection method can be combined with electrical spin rotation which smoothly leads to phase measurements of photoelectron spins.

In conclusion, we verified single photon to single electron spin angular momentum transfer in a gate defined DQD from single-shot measurements of the photoelectron spin states. Real-time signals of PSB enabled the detection of photon polarization states that are projected on the prepared electron spin. This result is an important step leading to photon to electron spin coherent conversion using gate defined lateral quantum dots in which spin-based quantum gate operation is feasible. The photon wavelength is selective by tuning the double heterostructure used for making QDs, and this enables resonant generation of single electron spins with single photons suitable for optical communication. The present result provides a useful scheme of hybrid photon-spin quantum systems with electrically-tunable solid state devices such as spin entanglement distribution which is essential for quantum communication.

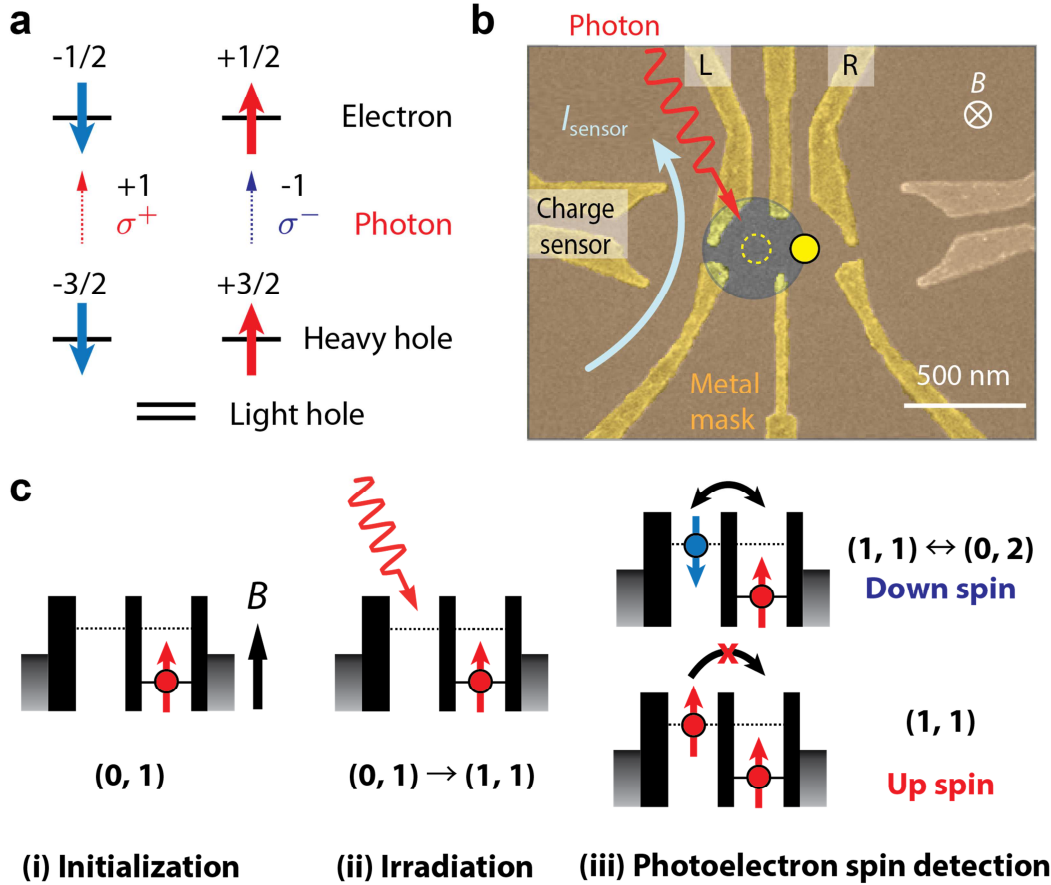

**Figure 1 | Photoelectron spin detection scheme. a**, Schematic of heavy hole band excitation in a GaAs QW used to selectively photoexcite an electron spin. The numbers represent the angular momenta of the respective particles. **b**, Picture of the sample overlaid with the shape of a metal mask with an optical aperture opened on top of the left QD. The solid circle represents the electron where one charge is occupied on the right dot before irradiation. Photoexcitation of the heavy hole band creates an electron-hole pair and the electrons are trapped in the left QD while holes are attracted to the surrounding gate electrodes. The charge sensor on the left is used to measure the electron tunnelling dynamics in real-time. **c**, Schematic of the photoelectron spin measurement procedure. A single up-spin is initially prepared on the right dot (i). Note that the up-spin state is the ground state of split Zeeman sublevels. After irradiation (ii), once an electron is trapped in the left QD (iii), the spin orientation is judged from a single-shot measurement of the PSB. If the photoelectron spin is down, the two-electron charge state undergoes a repeated transition between (1,1) and (0,2) states, whereas if the spin is up, no charge transition occurs. Finally the DQD is initialized to the (0,1) state by the escape of an excess electron to either lead.

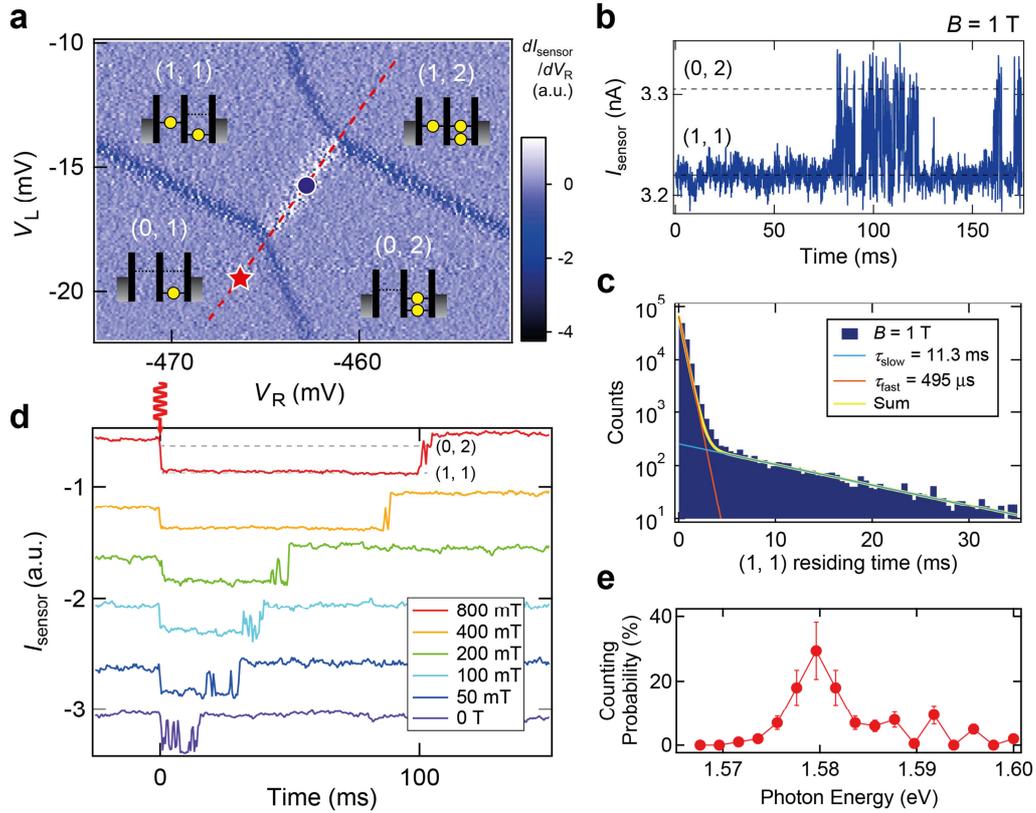

**Figure 2 | Real-time detection of PSB and single-shot photoelectron trapping in a DQD.**
**a**, Stability diagram measured in the vicinity of the inter-dot charge transition line (in white) between (1,1) and (0,2) charge states at a zero dot bias voltage. The white line indicates the resonance of the two states. **b**, Real-time trace of the charge sensor current $I_{sensor}$ taken at the resonance of (1,1) and (0,2) states (point ● in **a**) at $B = 1$ T showing the tunnelling features of PSB. Parallel and anti-parallel spin configurations are resolved as the time blocked in the (1,1) state and the time tunnelling back and forth between (1,1) and (0,2) states, respectively. **c**, Histogram of the (1,1) state residing time taken at $B = 1$ T showing a double exponential feature. Two decay times are obtained $\tau_{slow} = 11.3$ ms and $\tau_{fast} = 496$ μs each representing the lifetime of the parallel spin configuration and the tunnelling time of the anti-parallel spin configuration. **d**, Typical real-time traces of $I_{sensor}$ for a single-shot photoirradiation measurement when a parallel spin state is measured in a finite magnetic field. For the traces over 50 mT, the photo-generated state starts with a longer residing time at the (1,1) state. The residing time or lifetime of the parallel spin state becomes longer for a larger magnetic field, because PSB is better established. Linearly polarised light is irradiated for these measurements which can also generate anti-parallel spin states. **e**, Probability of detecting a photoelectron trap in various wavelength around the heavy hole band absorption peak.

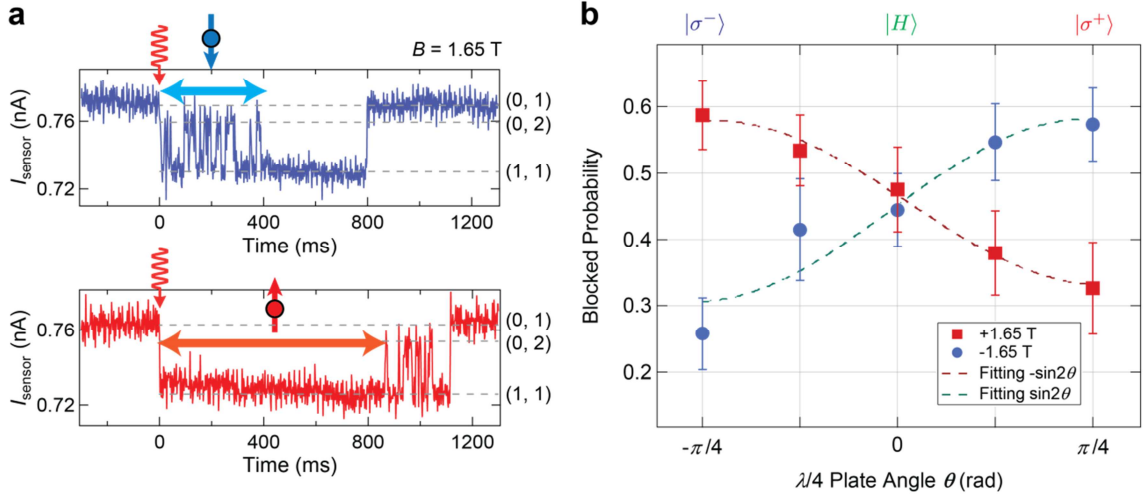

**Figure 3 | Single photoelectron spin detection and angular momentum transfer. a**, Typical real-time $I_{sensor}$ traces measured at ★ in Fig. 2a for single photoelectron trapping at $B$ = 1.65 T. Linearly polarized laser pulses create a (1,1) state having either anti-parallel or parallel spin with equal probability. The observed photoelectron is inferred to have down spin from the subsequent resonant tunnelling after irradiation (upper panel) and up spin from the initial blocked signal subsequent to irradiation (lower panel). **b**, Polarization dependence of the probabilities of detecting a parallel spin creation upon photoelectron trapping measured at ±1.65 T. The data points are fitted with a sine function indicating successful angular momentum transfer from circular polarization to electron spins with a heavy hole band excitation. Red dots for the positive magnetic field and blue dots for the negative field are mirrored, because the initially prepared electron has opposite spin between +1.65 T and −1.65 T. The error bars are calculated as the standard deviation of the binomial distribution of the detection outcome.

Methods summary

The irradiated dot is laterally defined by confining electrons of a two-dimensional electron layer formed in a 7.3 nm thick GaAs quantum well sandwiched by two $Al_{0.34}Ga_{0.66}As$ barriers grown on a (001) GaAs substrate. The wafer is the same as used in Ref. 19. With respect to the incident photon bandwidth of 600 μeV, the heavy hole band is well separated from the light hole band by confinement. The electron excitation spectrum of a gate defined QD follows the quantum well structure having a peak at 1.579 eV (785 nm wavelength) [19]. The single electron spin splitting in a QD with this specific QW has been investigated separately, giving an out-of-plane g-factor of 0.12 [28]. This corresponds to a Zeeman energy of 11.3 μeV at 1.65 T which is well within the laser bandwidth. Therefore the two spin states are treated equally with a single centre frequency photon. A 30 × 30 μm$^2$ wide and 200 nm thick Ti/Au metal layer is fabricated above the dot with a 400 nm diameter optical window to selectively irradiate the left QD using a laser beam with 7 μm diameter at the sample surface. The photons are picked from a Ti:Sapphire pulsed laser and focused using a commercial lens close to the sample on the cold finger. More information on the optics set-up is described in the Supplementary material.

Acknowledgements

We acknowledge T. Asayama for his contribution to the initial stage of this work. This work was supported by Grants-in-Aid for Scientific Research A (No. 25246005) and S (26220710), Innovative Areas "Quantum Cybernetics" (Grant No. 21102003) and "Nano Spin Conversion Science" (Grant No. 26103004), ImPACT Program of Council for Science, Technology and Innovation (Cabinet Office, Government of Japan), Intelligence Advanced Research Projects Activity project Multi-Qubit Coherent Operations through Copenhagen University, MEXT Project for Developing Innovation Systems, and QPEC, The University of Tokyo, The Strategic Information and Communications R&D Promotion Programme (SCOPE) of the Ministry of Internal Affairs and Communications Government of Japan (MIC). T. F. & H. K. were supported by JSPS Research Fellowships for Young Scientists. T. F. is supported by Yukawa Memorial Foundation (Mochizuki Fund). M. L. acknowledges support as an "International Research Fellow of the Japan Society for the Promotion of Science". A. L. and A. D. W. acknowledge gratefully support from DFG-TRR 160, Mercur Pr-2013-0001, BMBF -Q.com-H 16KIS0109, and DFH/UFA CDFA-05-06.

SUPPLEMENTARY INFORMATION

### A. Optical set-up and heavy hole excitation

We used a Ti:sapphire laser with a pulse picker to irradiate a train of pulses on the QD (Fig. S1a). The number of pulses included in this pulse train enables the tuning of the number of photons arriving on the QD. Additionally, two mechanical shutters helped to block the remaining scattered pulses. The shutters open for a time window of ~30 μs, which is shorter than the rise time of the charge sensor.

Photon polarizing components were aligned as follows. We set the laser beam direction and external magnetic field axis parallel to each other (z-axis) and perpendicular to the plane (x-y plane) of the DQD. The spins and circular polarizations are therefore both quantized along the z-axis which is toward the sample surface. We defined the positive $B$ field to favour electron up-spin with positive angular momentum (always in units of $\hbar$) of +1/2 and the clockwise (anti-clockwise) photon polarization σ+ (σ-) with the positive angular momentum of +1 (-1) both along +z.

The correspondence that we verify between circular polarization and the excited electron spin for the heavy hole excitation is obtained from the law of conservation of angular momentum. The heavy hole states have angular momentum of $\pm 3/2$, which relates the orbital angular momentum $\pm 1$ and the spin angular momentum $\pm 1/2$. A circular polarized photon, having angular momentum of $\pm 1$, interacts with the orbital motion of the electron and creates a conduction band electron having the orbital angular momentum of 0. From the relation between the orbital and spin angular momentum of the heavy hole, we obtain a polarization to spin relation that an angular momentum +1/2 electron spin (up spin) is excited with a -1 angular momentum photon (σ-) and an angular momentum -1/2 (down spin) is excited with a +1 angular momentum photon (σ-) [1].

### B. Single photoelectron trapping

Before starting the irradiation experiment, we have examined the relevant tunnelling rates to follow the photoelectron dynamics in a non-equilibrium DQD state. The inter-dot electron tunnelling rate is tuned slower than the measurement bandwidth of 10 kHz and the rates to the source and drain barriers even slower, in the order of 1 Hz. Especially the left barrier to the lead is raised so that the (1,1) state does not become a (0,1) state faster than the possible inter-dot oscillation between (1,1) and (0,2) states.

For irradiation measurements, the dot is initialized to the (0,1) state (point ★ in Fig. 2a) along the red dashed line with the (1,1) and (0,2) excited state levels aligned. Figure S1b shows typical real-time traces of laser pulse irradiation measurements with and without photoelectron trapping in zero magnetic field. These examples are taken with linearly polarized photons to generate both parallel and anti-parallel spin states by trapping a photoelectron. Although both spin configurations should be created, due to the mixing by hyperfine interaction we observe a single type of data. The trapped photoelectron resonantly tunnels back and forth between the two dots and then tunnels out to the lead (Fig. S1b top). The case where no electron is trapped (Fig. S1b bottom) is shown for reference where persistent photoconductivity acts as a small offset on the real-time traces.

### C. Distinguishability of real-time spin blockade

Here we discuss the distinguishability of the single-shot measurement by the PSB which can be measured prior to the photon irradiation measurement. In the spin measurement, parallel and anti-parallel spin configurations are distinguished by the relatively long and short tunnelling times observed in the real-time traces of the electron tunnelling events. These two time scales are numerically obtained by plotting a double exponential histogram of the (1,1) state residing time. A large number of data points are needed to clearly resolve these two decay times so the analysis of the PSB related lifetimes are performed on the measurements in the dark (for example Fig. 2b and 2c in the main text). We add an explanation in the next section on how the blockade lifetime in irradiation measurements relate to the PSB decay using coarse statistics.

The detection fidelity can be degraded by errors such as low signal-to-noise ratio of the charge sensor, missing detections due to the measurement bandwidth and the sorting errors that give the wrong decision of the measured spin due to the tunnelling time statistics that we focus in this section. We omit the discussion for the first two cases owing to the large signal-to-noise ratio over 5 and the long PSB lifetimes in the order of tens of milliseconds compared to the detector rise time of 100 μs, respectively.

The statistics of the residing time is extracted at a condition where we observe bi-stable sensor outcomes measured on the (1,1)-(0,2) states on resonance (Fig. 2b in the main text). This gives a clear double exponential decay (Fig. 2c in the main text). Therefore the tunnelling time statistics are simply represented by two decay time constants $\tau_{\text{slow}}$ and $\tau_{\text{fast}}$, which we attribute to a parallel spin lifetime and an anti-parallel spin tunnelling time, respectively.

We calculate the best threshold time $t_{th}$ that gives the least probability of errors. Given the two decay times, the error probability of sorting the slow and fast lifetimes are calculated with,

$$A \int_0^{t_{th}} e^{-t/\tau_{slow}} dt \quad \text{and} \quad B \int_{t_{th}}^{\infty} e^{-t/\tau_{fast}} dt,$$

where $A$ and $B$ are the expected probability of the histogram when t → 0 for the two decay times. We choose a specific condition for the calculation where the original spins are equally distributed ($A\tau_{slow} = B\tau_{fast} = 1/2$) and apply to the condition detecting unknown spins. Then the best $t_{th}$ is at the minimal value of the sum of the error rates which gives,

$$t_{th} = \frac{\tau_{slow}\tau_{fast}}{\tau_{slow} - \tau_{fast}} \ln\left(\frac{\tau_{slow}}{\tau_{fast}}\right).$$

We can calculate the expected error rates of the single-shot spin measurement by inserting this threshold value back into the formula of the error rates. Then we obtain the probability of judging the anti-parallel spin state for the parallel spin state by error as,

$$A\tau_{slow}\left(1 - r^{\frac{1}{1-r}}\right),$$

where $r = \frac{\tau_{slow}}{\tau_{fast}}$ and vice versa as

$$B\tau_{fast} r^{\frac{r}{1-r}}.$$

$A\tau_{slow}$ and $B\tau_{fast}$ correspond to the area of each exponential decay, meaning the probabilities of the original spin distribution.

We can use these formulae to estimate the threshold time and expected error rates of the photoelectron spin measurements. For example we derive $t_{th} = 67$ ms for decay times $\tau_{slow} \sim 500$ ms and $\tau_{fast} \sim 20$ ms obtained at $B = 1.65$ T. The probability of detecting a parallel spin correctly for a perfect parallel spin initialization would be 87 % and detecting an anti-parallel spin with its perfect initialization is 97 %. We see from these populations that although the detection is perfect there would be a shift towards less probability of detecting parallel spins states.

### D. Spin blockade of photoelectron spins

Here we explain how we attribute the origin of photoelectron PSB lifetimes such as those shown for the different fields in Fig. 2e in the main text. We observed two different regimes of the PSB lifetime for the *B* field below and over 400 mT. In the low field region, a monotonic increase of the lifetime time was easily observed in the real-time measurements as in Fig. 2c. This increase is compared with the experiment of Ref. 2, where the hyperfine interaction dominates the spin flipping and therefore the PSB lifetime. Note the upper bound of the low *B* region is significantly higher than that in Ref. 2. Our QW has a g-factor smaller than the bulk GaAs g-factor by a factor of 3~4. Therefore we need a higher magnetic field to decouple the parallel spin states from the nuclear spin field fluctuation.

For the higher fields, spin-orbit interaction is presumed to be dominant and in Ref. 3 they observe a spin flipped tunnelling with photon assisted tunnelling that is enhanced by spin-orbit coupling, which has no explicit dependence on *B* field. Experimentally we traced the *B* field dependence of the photoelectron PSB lifetime to qualitatively compare the dependence and ensure that the PSB mechanism is working in higher fields. We took statistics of the photoelectron PSB time and plotted for different *B* fields (Fig. S2). The histograms are plotted only for the tunnelling times of tens to hundred milliseconds to focus on the small number of acquired data. Each trace shows an exponential decay with no strong *B* field dependence indicating the presence of PSB.

### E. Efficiency of photoelectron spin detection

Here we discuss reasons for the degradation of angular momentum transfer in Fig. 3 in the main text, mainly focusing on the ~60 % probability of detecting a parallel spin state with σ- irradiation. We divide the problem to the spin detection error coming from the PSB distinguishability, spin initialization error, and the conversion error from photon to spin. The first error is discussed in section B giving a 13 % error rate.

The spin initialization would be degraded by the surrounding phonon environment. An up electron spin state can be excited to the down spin with a combination of phonons and spin-orbit interaction. In irradiation measurements, the interval from one photoelectron trapping event to the next is in the order of minutes so the spin has long enough time to fully interact with the environment. As a consequence, randomness is caused in the spin measurement axis, thus the degradation of photoelectron spin measurements. In Ref. 4 the authors argue that an electron spin in a QD is thermalized with a Boltzmann distribution. We estimate the phonon energy distribution from the average electron temperature of 100 mK determined at the inter-dot transition line from the stability diagram (Fig. 2a in the main text). If the

Zeeman split spin state is thermalized with this temperature, an error rate of 21 % is added in the case of $B = 1.65$ T.

      Taking into account these errors, originating from the electrical point of view of the spin measurement, the remaining error that come from the photoexcitation must be below 5~10 %. A few reasons can account for the errors such as polarization distortion at the metal gates, heavy hole-light hole mixing in the QW, and fine structure splitting by direct excitation of the QD. More detailed measurement and analysis are required to reveal the limiting factors of the photoelectron spin measurement.

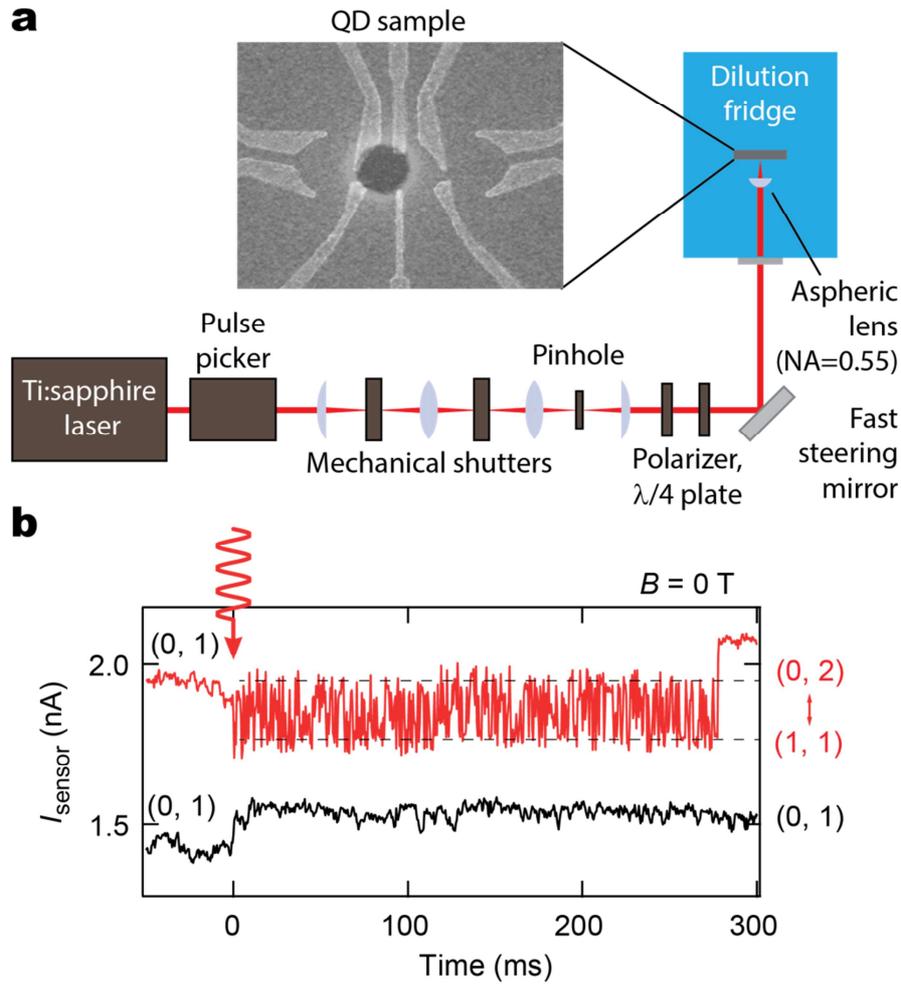

**Figure S1 | a**, Schematic of the optical set-up used for irradiating polarized photons to the sample in a single-shot manner. The pulse picker, mechanical shutters, and the voltage meter are synchronized by an external voltage pulse to obtain real-time signals. A fast steering mirror is used to scan the laser and find the centre position of the sample. The laser transmits through a transparent window on the cryostat and focused with a lens which is placed close to the sample on the cold finger. A scanning electron micrograph image of the gate patters of the DQD and the transparent image of the metal mask with an aperture is shown together in the figure. **b**, Typical real-time traces of $I_{sensor}$ for a single-shot photoirradiation measurement at zero. (upper trace) Trace showing a single photoelectron trapping where $I_{sensor}$ oscillates upon pulse irradiation due to a single photoelectron repeatedly tunnelling between the dots. The tunnelling times of the source and drain barriers are set to be long enough that the photo-generated two-electron state stays as a non-equilibrium state but to be detected in our measurement time scale. (lower trace) Trace without showing a photoelectron trapping indicating an offset caused by the irradiation.

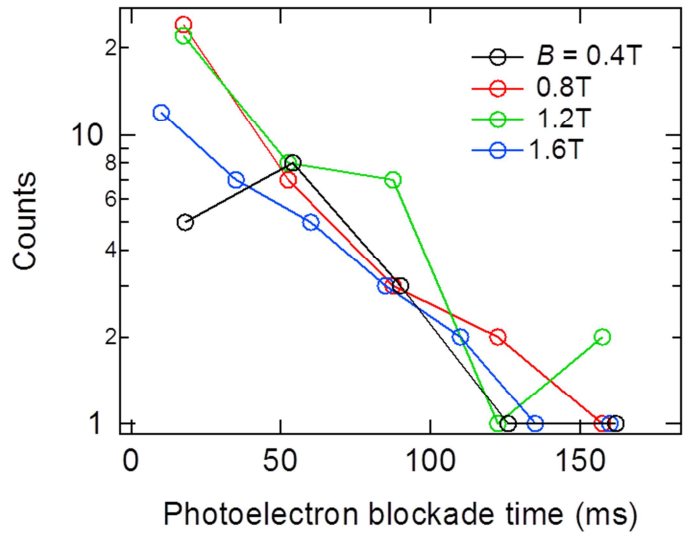

**Figure S2|** Histograms of the first (1,1) state residing time after photoelectron trapping at different fields over $B = 400$ mT. Data are plotted between tens and a hundred milliseconds that represent the photoelectron PSB lifetimes. Each histogram shows an exponential decay with a similar decay time, manifesting the decay to be dominated by spin-orbit interaction.